\documentclass[pra,twocolumn,floatfix]{revtex4}

\usepackage{graphicx}
\usepackage{color}
\usepackage[normalem]{ulem}
\usepackage{braket}

\newcommand{\be}{\begin{equation}}
\newcommand{\ee}{\end{equation}}
\newcommand{\ba}{\begin{eqnarray}}
\newcommand{\ea}{\end{eqnarray}}
\newcommand{\nn}{\nonumber \\}

\begin{document}

\title{Quantum eigenstate tomography\\ with qubit tunneling spectroscopy}

\author{Anatoly Yu. Smirnov and Mohammad H. Amin}
\affiliation{D-Wave Systems Inc., 3033 Beta Avenue, Burnaby BC
Canada V5G 4M9}
\begin{abstract}
Measurement of the energy eigenvalues (spectrum) of a multi-qubit
system has recently become possible by qubit tunneling spectroscopy
(QTS). In the standard QTS experiments, an incoherent  probe qubit
is strongly coupled to one of the qubits of the system in such a way
that its incoherent tunneling rate provides information about the
energy eigenvalues of the original (source) system. In this paper,
we generalize QTS by coupling the probe qubit to many source qubits.
We show that by properly choosing the couplings, one can perform
projective measurements of the source system energy eigenstates in
an arbitrary basis, thus performing quantum eigenstate tomography.
As a practical example of a limited tomography, we apply our scheme
to probe the eigenstates of a kink in a frustrated transverse Ising
chain.
\end{abstract}


 \maketitle

\section{Introduction}

Superconducting qubits are used as the basic building blocks for
practical implementation of scalable quantum computers~\cite{You05}.
In particular, the existing annealing-based quantum processing units
(QPU) \cite{Johnson11} exploit flux qubits based on superconducting
quantum devices (rf-SQUIDs)~\cite{Harris1Q, Harris8Q}. The qubits
are controlled by a limited number of low-bandwidth external lines.
This feature allows creating a quantum processor with more than 1000
qubits, while at the same time keeping a low level of noise in the
system. Experimental technique, termed  \textit{qubit tunneling
spectroscopy} (QTS) \cite{Berkley13}, has been developed in order to
measure quantum spectra of superconducting qubits (\textit{source}
qubits) using \textit{probe} qubits undergoing incoherent tunneling
transitions. A similar idea of weakly coupling a probe qubit to the
quantum system with the goal of observing its energy spectrum was
proposed in Ref.~\cite{Wang12}. In Ref.~\cite{Berkley13}, however,
the coupling between the probe qubit and the source qubits is strong
and the method was experimentally implemented with rf-SQUID flux
qubits. Quantum spectra, characterized by line splittings of the
order of few GHz, were measured using MHz-range control lines. The
same technique was also employed to demonstrate quantum entanglement
in systems of two and eight flux qubits embedded into an
industrial-scale quantum annealing processor~\cite{Lanting14}. These
experiments are performed with a unit cell having a linear size of
the order of 0.3~mm. Quantum spectra taken in the process show very
well resolved spectral lines, thus demonstrating a clear example of
macroscopic quantum coherence \cite{Leggett80} in a multi-qubit
system.

In QTS, information about quantum properties of the source qubits is
extracted from the bias dependence of the incoherent tunneling rate,
$\Gamma(\epsilon)$, of the probe qubit. Here $\epsilon$ is the
external flux applied to the probe along $z$-axis. Positions of the
maxima of $\Gamma(\epsilon)$ determine the energy levels of the
source system, whereas its peak amplitude is proportional to the
overlap between the eigenfunctions of the total (probe plus source)
system before and after the tunneling \cite{Wang12, Berkley13}.

In this paper, we generalize QTS technique to the case where the
probe qubit is coupled to many source qubits in an arbitrary basis.
 We show that projective measurements of energy eigenstates of the source system
 in an arbitrary basis can be performed with this approach. Therefore,
 our measurement scheme is tomgraphically complete. As a practical
 example, we consider dynamics of a kink in a frustrated
transverse Ising chain, in which the nearby qubits are coupled
ferromagnetically and the first and the last qubits are biased in
the opposite direction. The classical states with the lowest energy
have a kink, which can be located between any nearby qubits. This
kink behaves like a free particle confined in a potential well. We
provide numerical calculations of the incoherent tunneling rate of
the system taking into account the low frequency environment. The
maxima of the tunneling rate, plotted as a function of the bias
applied to the probe, are shown to be proportional to the modulus
squared of the eigenfunctions.

\section{Qubit tunneling spectroscopy}

Following Ref.~\cite{Berkley13}, here and in Appendix~\ref{QTS} we
derive  a set of formulas describing multi-qubit QTS experiments.
The quantum system under study has $N$ coupled qubits, with
Hamiltonian $H_S$ written in terms of of Pauli matrices $\{
\sigma^x_i, \sigma^y_i, \sigma^z_i\}$ where $i \in \{1, \ldots,N\}$.
We denote eigenstates and eigenvalues of this Hamiltonian by
$\ket{\Psi_n}$ and $E_n$, respectively, where $n \in \{0,
1,\ldots,(2^N-1)\}$. There is no need to specify Hamiltonian $H_S$
of the quantum system at this stage, although later we will consider
the transversal Ising chain as an example. In addition to the above
{\em source} qubits operating in a fully quantum regime, we have one
\textit{probe} qubit characterized by a small tunneling amplitude
$\Delta_p$ and by an external bias $\epsilon$. This qubit works in
an incoherent regime of macroscopic resonant tunneling (MRT)
\cite{AminAverin08, Harris08}. We write the total source-probe
Hamiltonian as:
\ba \label{Ham1} H_0 = H_S + (H_C + \epsilon/2) (1 - \sigma^z_p ) -
\Delta_p \sigma^x_p \ea
The probe qubit, which is described by the set of its Pauli matrices
$\{\sigma^x_p, \sigma^y_p, \sigma^z_p\},$ has two eigenstates,
$\ket{\uparrow_p}$ and $\ket{\downarrow_p},$ of the matrix
$\sigma^z_p: \;\sigma^z_p \ket{\uparrow_p} = \ket{\uparrow_p}, \;
\sigma^z_p \ket{\downarrow_p} = - \ket{\downarrow_p}.$ The coupling
between the source qubits and the probe is provided by the term: $-
\sigma^z_p H_C$. Once again, the details of $H_C$ does not affect
our general description. The second term in the Hamiltonian vanishes
when probe qubit is in state $\ket{\uparrow_p}$. Therefore, we can
write
\ba \label{Ham2} H_0 = H_S^\uparrow \otimes
\ket{\uparrow_p}\bra{\uparrow_p} + H_S^\downarrow \otimes
\ket{\downarrow_p}\bra{\downarrow_p} - \Delta_p \sigma^x_p, \ea
where $H_S^{\uparrow} = H_S$ and $H_S^{\downarrow} =
H_S+2H_C+\epsilon$. We denote the eigenstates of these Hamiltonians
by $\ket{\Psi_n^{\uparrow}}$ and $\ket{\Psi_m^{\downarrow}}$:
 \ba \label{Psi1}
 H_S\ket{\Psi_n^{\uparrow}} &=& E_n^\uparrow \ket{\Psi_n^{\uparrow}}, \nn
 H_S^\downarrow \ket{\Psi_m^{\downarrow}} &=& (E_m^\downarrow + \epsilon) \ket{\Psi_m^{\downarrow}}.
 \ea
Notice that $\ket{\Psi_n^{\uparrow}}=\ket{\Psi_n}$ and $E_n^\uparrow
= E_n$.

In the limit of small $\Delta_p$, the eigenstates of $H_0$ are
approximately $\ket{\psi_n^{\uparrow}} =
\ket{\Psi_n^{\uparrow}}\otimes\ket{\uparrow_p}$ and
$\ket{\psi_m^{\downarrow}}=\ket{\Psi_m^{\downarrow}}\otimes\ket{\downarrow_p}$,
where $n,m\in\{0,1,\ldots,(2^N-1)\}$. When the probe qubit is in its
``up" state, $\ket{\uparrow_p}$, the eigenstates of the source
qubits coincide with the eigenstates of the original Hamiltonian
$H_S$. Notice that this is true even when coupling to the probe
qubit is strong. In the opposite case, when the probe is in its
``down" state, $\ket{\downarrow_p}$, the coupling to the probe
should create a large bias for the source qubits in such a way that
the new Hamiltonian, $H_S^{\downarrow}$, has a well-defined ground
state $\ket{\Psi_0^{\downarrow}}$ isolated from the rest of the
eigenstates by a large energy gap. The total system is then
initialized in $\ket{\psi_0^\downarrow } =
\ket{\Psi_0^\downarrow}\otimes\ket{\downarrow_p}$ and tunnels to one
of the eigenstates $\ket{\psi_n^\uparrow } =
\ket{\Psi_n^\uparrow}\otimes\ket{\uparrow_p}$, as the probe tunnels
from  $\ket{\downarrow_p}$ to $\ket{\uparrow_p}$. The two sets of
states of the total Hamiltonian $H_0$, with the probe qubit being in
$\ket{\downarrow_p}$ or $\ket{\uparrow_p}$, can be  moved relative
to each other by applying a probe bias $\epsilon$. The rate of
macroscopic tunneling between the initial and final states can show
well-resolved peaks when the eigenenergies of these states are in
resonance \cite{Harris08}.

To have a full description of the dynamics we consider the system
being exposed to an environment. As shown in Ref.~\cite{Lanting14},
the width of the MRT peaks is predominantly determined by the low
frequency noise coupled to the probe qubit. As such, in our modeling
we consider an environment only interacting with the probe qubit.
The dissipative dynamics of the probe-source system is therefore
described by the master equation (\ref{P3}) written in our case as
(see Appendix \ref{QTS})
\ba \label{P4} \dot P_0 + \Gamma_0 P_0 = \sum_n \Gamma_{0n} P_n, \ea
where $P_0$ and $P_n$ are the probabilities of being in state
$\ket{\psi_0^\downarrow }$ and   $\ket{\psi_n^\uparrow }$,
respectively, and
\ba \label{Gam5} \Gamma_{0}(\epsilon) =\Delta_p^2\, \sum_n |\langle
\Psi_n^\uparrow | \Psi_0^\downarrow \rangle |^2 \times \nn
\sqrt{\frac{2 \pi}{ W^2}}\; \exp \left[ - \frac{(E_n^\uparrow -
E_0^\downarrow - \epsilon + \epsilon_p)^2 }{2\, W^2 } \right], \\
\label{Gam6} \Gamma_{0 n}(\epsilon) = \Delta_p^2\, |\langle
\Psi_n^\uparrow | \Psi_0^\downarrow \rangle |^2 \times \nn
\sqrt{\frac{2 \pi}{ W^2}}\; \exp \left[ - \frac{(E_n^\uparrow -
E_0^\downarrow - \epsilon - \epsilon_p)^2 }{2\, W^2 } \right]. \ea
The MRT width $W$ and the reorganization energy $\epsilon_p$ are
related by fluctuation dissipation theorem (see
Ref.~\cite{AminAverin08} and Appendix \ref{QTS}). At $t=0$, we have
$P_n=0$, and therefore the initial slope of probability decay is
only given by $\Gamma_0$, which is measured in the MRT experiments.
At the point of resonance, where $\epsilon = E_n^\uparrow -
E_0^\downarrow + \epsilon_p$, the peak value of the escape rate
$\Gamma_0$ is proportional to the overlap of the initial and final
wave functions,
\ba \Gamma^{\rm peak} \propto |\langle \Psi_n^\uparrow |
\Psi_0^\downarrow \rangle |^2. \ea
In QTS, the probe qubit bias, $\epsilon$, is swept within some range
and $\Gamma_0(\epsilon)$ is measured. The locations of the peaks of
$\Gamma_0$ give information about the energy spectrum, $E_n$, of the
source system.

\section{Quantum eigenstate tomography}

For spectroscopy purposes, the details of $H_C$ are unimportant as
long as the ground state of $H_S^\downarrow$ is well separated from
excited states and has overlap with the eigenstates of
$H_S^\uparrow$ ($=H_S$) that we want to detect. Since the tunneling
rate is proportional to
$|\langle\Psi_n^{\uparrow}|\Psi_0^{\downarrow}\rangle|^2$, if we can
set $\ket{\Psi_0^{\downarrow}}$ as an arbitrary state, we can
measure the projection of $\ket{\Psi_n^{\uparrow}}$ on that state
and therefore perform tomography on the source system's eigenstates,
$\ket{\Psi_n}$. This can be done by choosing a specific form of
$H_C$. For example, assume that the probe qubit can be coupled to
\textit{all} source qubits along \textit{all} three axes. We can
therefore write
 \be \label{Ham3}
 H_C = \sum_{i=1}^N ( J_{pi}^x \,\sigma^x_i + J_{pi}^y\, \sigma^y_i + J_{pi}^z \,\sigma^z_i )
 \ee
The coupling constants $J^x_{pi}, J^y_{pi}$ and $J^z_{pi}$ create
additional biases for the $i$-th source qubit along $x$, $y$, and
$z$ directions. These biases disappear in the case when the probe
qubit is in $\ket{\uparrow_p}$ and, as before, $H_S^\uparrow=H_S$.
In the opposite direction of the probe, $\ket{\downarrow_p},$ strong
coupling constants $\{J^x_{pi}, J^y_{pi}, J^z_{pi}\}$ can make $H_C$
dominate in $H_0$, thus suppressing the contribution of $H_S$. The
standard $z$-$z$ couplers \cite{Harris1Q, Harris8Q}  between the
probe and source qubits lead to the following set of the constants:
$\{ 0, 0, J^z_{pi} \}$. This set creates the $z$-directed initial
state $\ket{\Psi_0^{\downarrow}} = \otimes_{i=1}^N \ket{z_i}, $
defined in terms of the eigenstates $\ket{z_i}$ of the matrices
$\sigma^z_i.$ Again, here the probe qubit is in the
$\ket{\downarrow_p}$ state. The up or down direction of the specific
$i$-th source qubit in the reference state
$\ket{\Psi_0^{\downarrow}}$ depends on the sign of the corresponding
coupling coefficient $J_{pi}^z.$ If instead of $z$-$z$ couplers we
have $x$-$z$ couplers, i.e., $J^x_{pi} \neq 0$ and $J^y_{pi} =
J^z_{pi} = 0,$ then the \textit{reference} state will be
$\ket{\Psi_0^{\downarrow}} = \otimes_{i=1}^N \ket{x_i}$, where
$\ket{x_i}$ are the eigenfunctions of $\sigma^x_i$. We can therefore
project the eigenstates $\ket{\Psi_n}$ onto the $x$-basis. Likewise,
we can project $\ket{\Psi_n}$ onto $y$-basis. Being able to do
projective measurements in all basis makes our protocol
tomographically complete.

In practice, the coupling of the probe qubit to all source qubits in
all bases could be challenging. However, with a limited number of
couplers working in a single basis we can still do projective
measurements in a very limited Hilbert subspace and obtain useful
information. For example, in ferromagnetic systems, one can do
projective measurements of the eigenstates onto the lowest energy
subspace $\{\ket{\uparrow\uparrow\ldots\uparrow},
\ket{\downarrow\downarrow\ldots\downarrow} \}$ with one coupler
\cite{Lanting14}. In the next section, we provide another example in
which the coupling of the probe qubit to two source qubits is needed
for projection onto the lowest energy subspace.

\section{Wave function of a kink in the frustrated Ising chain}

\begin{figure}
\includegraphics[width=0.5\textwidth]{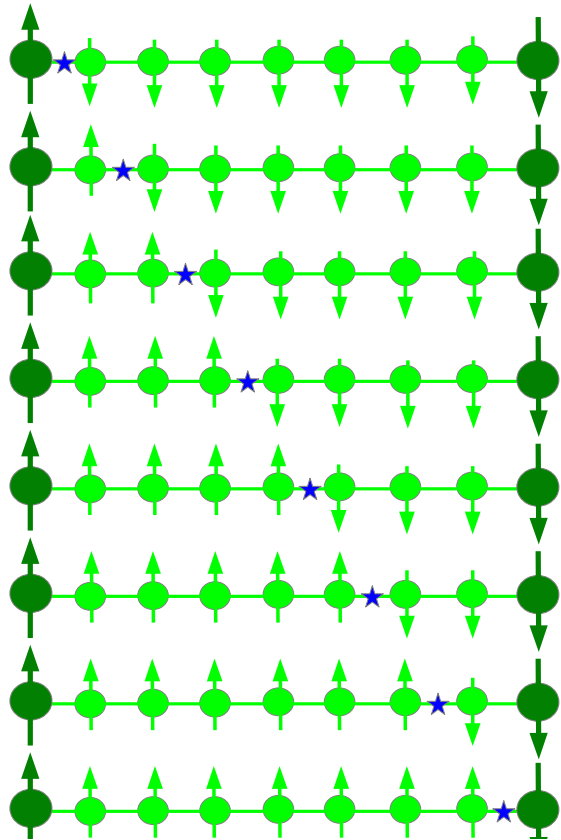}
\caption{\label{FIG:Chain1} Eight possible positions of a kink
(shown as a blue star) in a chain with seven source qubits and two
boundary spins. A left boundary qubit is always up, and a right
boundary qubit is always down.}
\end{figure}

As an example of the source quantum system we consider a frustrated
Ising chain described by the following Hamiltonian $H_S$,
\ba \label{Hs1} H_S = \sum_{i-1}^N ( h_i \, \sigma^z_i - \Delta_i\,
\sigma^x_i) + \sum_{i<j} J_{ij} \sigma^z_i \sigma^z_j. \ea
For this specific case the transverse field is determined by the set
of tunneling amplitudes $\{\Delta_i\}$. The qubits are biased along
$z$-direction with the strengths $\{h_i\}$ and coupled with the
constants $\{J_{ij}\}$. In principle, the qubits can be biased and
coupled along any direction, $x$, $y$, or $z$. The chain has $N$
qubits, with a uniform ferromagnetic coupling between the nearby
qubits, $J^z_{i j}  = - J\, \delta_{i\pm 1, j}, $ where $J
> 0.$ Notice that hereafter we work in the $z$-basis since the
probe qubit is coupled to $\sigma^z_i$-matrices of the source
qubits. This can be done by means of the standard magnetic couplers
\cite{Harris1Q, Harris8Q}.
 The frustration is created by two additional boundary
qubits, which have fixed spin directions. In Fig.~\ref{FIG:Chain1}
we show configurations of seven source spins ($N = 7$) plus two
boundary qubits.
The left boundary qubit is always in the up-direction, and the right
boundary qubit is always directed downwards. These qubits are
ferromagnetically coupled to the first and to the last  qubits in
the Ising chain, thus creating a bias $h_1 = -J$ applied to the
first qubit and the bias $h_N = J$ applied to the last qubit in the
chain. The other qubits have zero biases, $h_i = 0$ at $i =
2,\ldots, N-1.$ In the absence of the probe qubit the lowest energy
of the $N$-spin Ising chain is degenerate. All eigenstates of the
source Hamiltonian $H_S$ with the lowest energy have one kink
located between the source qubits. A kink also can be located
between the first qubit in the chain and the left boundary qubit,
and between the right boundary qubit and the last qubit in the
chain. The possible locations of the kink, which is shown as a blue
star, are presented in Fig.~\ref{FIG:Chain1}. It is known that the
kink in the Ising chain behaves like a free quantum particle. The
main goal of this part of the paper is to demonstrate that the QTS
measurements allow us to visualize the quantum distributions of the
kink in the frustrated Ising chain.

In the process of QTS tomography the lowest eigenstates of the
kink's Hamiltonian are projected on the basis set formed by vectors
that have  a definite kink location. In particular, functions
$\ket{\psi_1},\ldots,\ket{\psi_{N+1}},$ where
\ba \label{kinkB} \ket{\psi_1} &=& \ket{\downarrow_1,\downarrow_2,
\ldots, \downarrow_{N-1}, \downarrow_N}, \nn \ket{\psi_2} &=&
\ket{\uparrow_1,\downarrow_2, \ldots, \downarrow_{N-1},
\downarrow_N}, \ldots, \nn \ket{\psi_{N}} &=&
\ket{\uparrow_1,\uparrow_2, \ldots, \uparrow_{N-1},\downarrow_N},
\nn \ket{\psi_{N+1}} &=& \ket{\uparrow_1,\uparrow_2, \ldots,
\uparrow_{N-1},\uparrow_N}, \ea
corresponds to $N+1$ positions of the kink in the frustrated Ising
chain (see Fig.~\ref{FIG:Chain1} for the case of seven source
qubits,  $N=7$). These functions form the quantum-mechanical basis.
We notice that the basis $\ket{\psi_l}$ is not complete since the
high-energy states with many kinks are neglected here. Every state
from the set  $\{\ket{\psi_1},\ldots,\ket{\psi_{N+1}}\}$ is
characterized by a definite position of the kink. An arbitrary
quantum state, for example, the $n$-eigenstate
 of the source
qubits, $\ket{\Psi_{n}^\uparrow} \equiv \ket{\Psi_{n}}$, can be
represented as a superposition of the basis states $\ket{\psi_l}$,
\begin{equation}
\ket{\Psi_{n}^\uparrow} = \sum_{l=1}^{N+1} \braket{\psi_l
|\Psi_{n}^\uparrow}\, \ket{\psi_l}.
\end{equation}
The set of amplitudes, $C_l^{(n)} = \braket{\psi_l
|\Psi_{n}^\uparrow}$, taken as functions of the quantum number $l$,
describes a wave function of the $n$-th energy state in a
single-kink representation. The quantum number $l$ serves as a
position of the kink in the frustrated Ising chain. Thus, the
$l$-dependence of the kink amplitude $C_l^{(n)}$ is equivalent to
the coordinate dependence of the wave function of the particle in
the state corresponding to the $n$-eigenstate of the source
Hamiltonian $H_S$. Here we have $n = 0, 1, \ldots, (2^N-1).$

\begin{figure}
\includegraphics[width=0.5\textwidth]{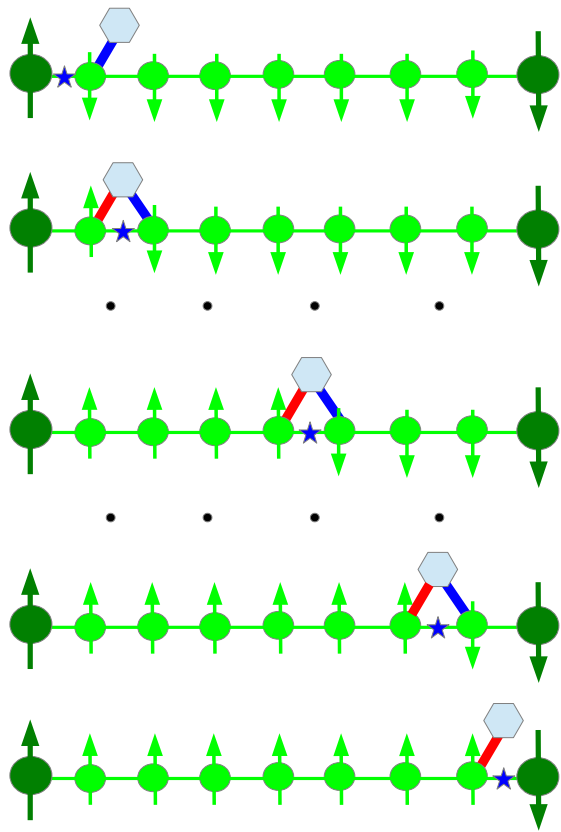}
\caption{\label{FIG:Chain2} A generation of basis states
$\ket{\psi_l}$ of the left manifold by selective coupling of the
probe qubit (shown as a hexagon) to the source qubits.  A positive
probe-source coupling $J_{pi}$ is drawn in blue, a negative $J_{pi}$
is shown in red. A blue star symbol is related to the kink location.
}
\end{figure}

We notice that the escape rate $\Gamma_0(\epsilon)$ (\ref{Gam5}) is
proportional to the overlap squared,  $| \langle \Psi^\uparrow_{n} |
\Psi^\downarrow_{0} \rangle |^2$,  of the $n$-eigenstate
$\ket{\Psi^\uparrow_{n}}$ of the source Hamiltonian $H_S$ and the
ground state $\ket{\Psi^\downarrow_{0}}$ of the biased source
qubits. The ground state $\ket{\Psi_0^\downarrow} $ of the left
manifold can be transformed into a specific basis state
$\ket{\psi_l} $, so that $\ket{\Psi_0^\downarrow} = \ket{\psi_l}$,
by choosing proper couplings $J_{pi}$ between the probe and the
source qubits as it is shown in Fig.~\ref{FIG:Chain2}. For example,
the first state $\ket{\psi_1} = \ket{\downarrow_1
\ldots\downarrow_N} $ can be generated if the first qubit in the
chain is coupled to the probe with a positive constant $J_{p1} = J_p
> 0.$ Other source qubits are decoupled from the probe. The second basis state $\ket{\psi_2} =
\ket{\uparrow_1 \downarrow_2 \ldots \downarrow_N}$ is created when
the probe qubit is coupled to the first source qubit by negative
coupling, $J_{p1} = - J_p,$ whereas its coupling to the second qubit
is positive, $J_{p2} = J_p > 0.$ To generate the state $\ket{\psi_l}
= \ket{\uparrow_1 \uparrow_2 \ldots \uparrow_{l-1} \downarrow_l
\ldots \downarrow_N}$, two nearby qubits $l-1$ and $l$ should be
coupled to the probe with opposite coupling strengths, $J_{p,l-1} =
-J_p$ and $J_{p,l} = J_p$. The last state, $\ket{\psi_{N+1}} =
\ket{\uparrow_1 \ldots \uparrow_N} $, is generated when the
$N$-qubit is coupled to the probe with a negative coupling strength,
$J_{pN} = J_p$, and other source qubits do not interact with the
probe.

\subsection{Quantum distribution of a kink}

\begin{figure}
\includegraphics[width=0.5\textwidth]{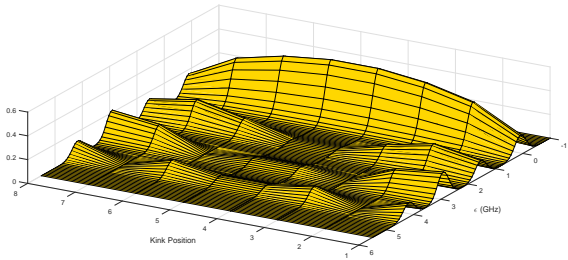}
\caption{\label{FIG:Gam7} Escape rate $\Gamma_0$ as a function of
the kink position $l$ and the bias $\epsilon$ applied to the probe
qubit. Here we have seven source qubits. As in the case of a free
particle in the potential well, the ground state of the kink has a
maximum in the middle of the chain where the first excited state has
a node. The second and third excited states have two and three
nodes, respectively.}
\end{figure}

\begin{figure}
\includegraphics[width=0.5\textwidth]{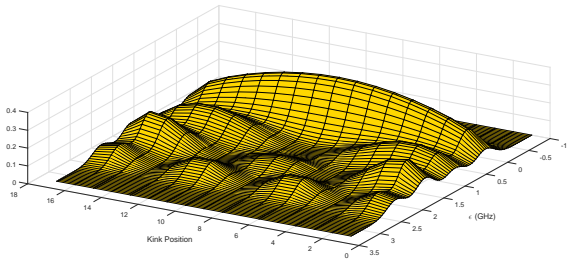}
\caption{\label{FIG:Gam16} the QTS rate $\Gamma_0(\epsilon,l)$
reflects the quantum distribution of a kink position $l$ in the
16-qubit Ising chain for four lowest energy eigenstates. The peak
with the energy $\epsilon \sim 0$ corresponds to the ground state of
the kink (particle). The next peak, with $\epsilon \sim 1$~GHz, is
related to the first excited state having one node, at $l = 8$, in
the wave function. The wave function of the next state, with the
energy $\epsilon\sim 1.5$~GHz, has two nodes, at $l=6$ and $l=12.$
The last state shown in the picture has the energy $\epsilon \sim
2.25$~GHz. This state is described by the wave function having three
nodes located at $l=4$, $l=9$, and $l=14.$ }
\end{figure}

In order to obtain the quantum distribution of the system over all
possible positions of the kink in the Ising chain we have to measure
the $l$-dependence of the function $|C_l^{(n)}|^2~=~|\braket{\psi_l
|\Psi_{n}}|^2$. To do that, we choose a specific connection between
the probe and source qubits related, for example, to the state
$\ket{\psi_l}$,  with a subsequent measurement of the escape rate
$\Gamma_0(\epsilon)$ for all possible probe biases $\epsilon$. As
the next step, we change the initial source-probe connection and
repeat the measurements. Finally, we obtain the rate
$\Gamma_0(\epsilon,l)$ as a function of the bias $\epsilon$ and the
kink position, which is characterized by the number $l$ of the
single-kink basis state $\ket{\psi_l}$. In Fig.~\ref{FIG:Gam7} we
show the normalized function $\Gamma_0(\epsilon,l)$ for the case of
seven qubits in the chain ($N = 7$).
Both figures are plotted at $\Delta_1 = \ldots = \Delta_N = 2$~GHz,
$J = 2$~GHz, $J_p = J$, with a temperature of $T = 12$~mK and a MRT
linewidth of $W = 10$~mK. Only four lowest energy levels of source
qubits, with $n=0,1,2,3$, are presented. In Fig.~\ref{FIG:Gam16} we
plot the QTS rate $\Gamma_0(\epsilon,l)$ for the chain that has 16
qubits ($N=16$) and for the same set of parameters. This figure has
a better resolution than Fig.~\ref{FIG:Gam7}.
In both figures, along the $\epsilon$-axis  we have the standard QTS
peaks corresponding to four energy eigenstates $\ket{\Psi_{n}}$ of
the source Hamiltonian $H_S$ (\ref{Hs1}). If we move along the other
axis, we will see a dependence of the states $\ket{\Psi_{n}}$ on the
kink position. The probability distributions of the kink in the
frustrated chain shown in Figs.~\ref{FIG:Gam7} and \ref{FIG:Gam16}
are similar to the distribution of a quantum particle in a potential
well.

\section{Conclusions}

We have generalized the qubit tunneling spectroscopy approach of
Ref.~\cite{Berkley13} to allow performing quantum eigenstate
tomography in a multi-qubit (source) system. An additional (probe)
qubit, working in the incoherent regime, has to be coupled to all
source qubits in all bases to make projective measurement onto an
arbitrary basis state possible. A limited, but practical, version of
tomography is described with an example of a single kink in a
frustrated Ising chain. The lowest energy eigenstates of this system
is equivalent to those of a free quantum particle confined in a
potential well. We have calculated the incoherent tunneling rate of
the system and shown that its peak values correspond to the modulus
squared of the overlap of the eigenstates and a preselected basis
state which is related to a kink position.

\section*{Acknowledgements}
 We are thankful to Professor Alexander Omelyanchouk for collaboration and support in early days of D-Wave Systems.
We are grateful to Chris Rich for helpful discussions and Fiona
Hanington for critical reading of the paper.

\appendix
\section{Derivation of the master equation}
\label{QTS}

The system of $N$ source qubits  and one probe qubit is described by
the Hamiltonian $H_0$  defined by Eqs.~(\ref{Ham1}) and
(\ref{Ham2}). The probe qubit is working in a regime of incoherent
tunneling between its wells. The tunneling is introduced by the
small tunneling amplitude $\Delta_p$ in the Hamiltonian $H_0$. We
also take into account an interaction of the probe qubit with its
dissipative environment, which is described by the variable $Q_p$.
Weak coupling of source qubits to their environments is omitted
here. This coupling contributes to the width of MRT lines of the
probe qubit\cite{Lanting14}. The main contribution to the linewidth,
however, is given by the low-frequency bath directly coupled to the
probe. The total Hamiltonian $H$ of the source-probe system coupled
to the probe qubit bath has the form
\ba \label{H1} H = H_0 - Q_p \,\sigma^z_p + H_B, \ea
where $H_B$ is the free Hamiltonian of the bath, and $\sigma^x_p$ is
the Pauli matrix responsible for the flipping of the probe qubits
between states $\ket{\downarrow_p}$ and $\ket{\uparrow_p}, $ $
\sigma^x_p = \ket{\uparrow_p}\bra{\downarrow_p} +
\ket{\downarrow_p}\bra{\uparrow_p}.$

The bath can be represented as a sum of independent harmonic
oscillators \cite{Mag59,Leggett87} with the Hamiltonian
\ba \label{HB} H_B = \sum_k \left( \frac{p_k^2}{2 m_k} + \frac{m_k
\omega_k^2 x_k^2}{2} \right). \ea
The $k$-th oscillator in the bath is characterized by  position
$x_k$,  momentum $p_k$, mass $m_k$ and  positive frequency $\omega_k
$. The bath operator $Q_p$ in Eq.~(\ref{H1}) is given by the formula
\ba Q_p = \sum_k m_k \omega_k^2 z_{kp} x_k. \ea
A constant $z_{kp}$ determines the strength of coupling between the
probe qubit and the $k$-mode of the bath. The Hamiltonian $H$
(\ref{H1}) can be written as
\ba \label{H2} H = H_S^\uparrow \otimes
\ket{\uparrow_p}\bra{\uparrow_p} + H_S^\downarrow \otimes
\ket{\downarrow_p}\bra{\downarrow_p}  - \Delta_p\, \sigma^x_p + \nn
\sum_k \frac{p_k^2}{2 m_k} + \sum_k \frac{m_k \omega_k^2}{2}\, (x_k
- z_{kp}\, \sigma^z_p )^2. \ea
Here we have omitted the constant term.
 The unitary
transformation
\ba U_p = e^{-i \xi_p\, \sigma^z_p} = \nn 1 + ( e^{- i \xi_p}  - 1
)\, \ket{\downarrow_p}\bra{\downarrow_p} + ( e^{ i \xi_p} - 1 )\,
\ket{\uparrow_p}\bra{\uparrow_p}, \ea
applied to the Hamiltonian $H$ turns this operator to the form
\ba \label{Ham5} H' = U_p^\dag H U_p = \sum_k \left( \frac{p_k^2}{2
m_k} + \frac{m_k \omega_k^2 x_k ^2}{2} \right) + \nn H_S^\uparrow
\otimes \ket{\uparrow_p}\bra{\uparrow_p} + H_S^\downarrow \otimes
\ket{\downarrow_p}\bra{\downarrow_p}  - \nn
 \Delta_p\, e^{2 i \xi_p}
\, \ket{\downarrow_p}\bra{\uparrow_p}  - \Delta_p\, e^{ - 2 i \xi_p}
\, \ket{\uparrow_p}\bra{\downarrow_p}. \ea
Here $\xi_p = \sum_k z_{kp}\, p_k$ is a stochastic phase produced by
the bath. We notice that this phase appears only at the tunneling
terms.

The source-probe Hamiltonian has eigenstates $\ket{\Psi_{\mu}} $
defined by the following equations:
\ba ( H_S^\uparrow \otimes \ket{\uparrow_p}\bra{\uparrow_p} +
H_S^\downarrow \otimes \ket{\downarrow_p}\bra{\downarrow_p} )
\ket{\Psi_{\mu} } = E_{\mu} \ket{\Psi_{\mu}}, \ea
where $\mu \in \{ 0,1, \ldots, (2^{N+1}-1)\}.$ We notice that the
set of eigenstates $\{\ket{\Psi_{\mu}}\}$ contains two subsets $-$
one, which is related to the up-state of the probe qubit, and
another, which is related to the down-state of the probe:
$\{\ket{\Psi_{\mu}}\} = \{ \ket{\Psi_n^\uparrow} \otimes
\ket{\uparrow_p}, \ket{\Psi_m^\downarrow} \otimes
\ket{\downarrow_p}$. Here the eigenstates $\ket{\Psi_n^\uparrow} $
and $\ket{\Psi_m^\downarrow} $ can be found from Eqs.~(\ref{Psi1}).
The indices $m$ and $n$ run over $2^N$ states: $m, n \in \{ 0,1,
\ldots, (2^N-1)\}.$ The eigenenergies $\{E_{\mu}\}$ also have two
subsets: $\{E_{\mu}\} = \{E_n^\uparrow, E_m^\downarrow +
\epsilon\},$ with $\epsilon$ being the bias applied to the probe
qubit.

Following the approach proposed in Ref.~\cite{ES81} and developed in
Ref.~\cite{SMN09} we introduce a time-dependent Heisenberg operator
$\rho_{\mu\nu}$ of the source-probe system, $$\rho_{\mu \nu} = (
\ket{\Psi_{\mu}}\bra{\Psi_{\nu} } )(t).$$ In the Heisenberg
representation the total Hamiltonian $H$ (\ref{Ham5}) is given by
the formula
\ba \label{Ham6} H = \sum_{\mu} E_{\mu} \, \rho_{\mu \mu} -
\sum_{\mu \neq \nu} Q_{\mu\nu} \, \rho_{\mu\nu} + H_B, \ea
with the bath operator
\ba \label{Q1}  Q_{\mu\nu} = \Delta_p\, e^{ 2 i \xi_p} \, \langle
\Psi_{\mu} | \downarrow_p\rangle \langle \uparrow_p | \Psi_{\nu}
\rangle \ + \nn \Delta_p\, e^{ -2 i \xi_p} \, \langle \Psi_{\mu} |
\uparrow_p\rangle \langle \downarrow_p | \Psi_{\nu} \rangle. \ea
Here we drop the prime sign in the Hamiltonian $H$ (\ref{Ham5}). The
diagonal elements of the Hamiltonian (\ref{Ham6}) are determined by
the eigenenergies $E_{\mu}$ of the system-probe Hamiltonian $H_0$
(\ref{Ham2}). The bath Hamiltonian $H_B$ is defined by
Eq.~(\ref{HB}). The operator $\rho_{\mu\nu}$ obeys the Heisenberg
equation
\ba \dot{\rho}_{\mu\nu} = i\,\omega_{\mu\nu} \, \rho_{\mu\nu} + i\,
\sum_{\mu'} ( Q_{\nu\mu'} \rho_{\mu\mu'} - Q_{\mu' \mu}
\rho_{\mu'\nu} ), \ea
with $\omega_{\mu\nu} = E_{\mu} - E_{\nu}.$ Using the approach
developed in Refs.~\cite{ES81,SMN09} we derive a set of equations
for the qubit operators averaged over fluctuations of the free bath,
\ba \label{NME} \braket{\dot{\rho}_{\mu\nu}} - i\,\omega_{\mu\nu} \,
\braket{\rho_{\mu\nu}} = \nn - \int_0^t dt_1
\braket{Q^{(0)}_{\nu\mu'}(t) Q^{(0)}_{\mu''\nu''}(t_1) }\, \braket{
\rho_{\mu\mu'}(t) \rho_{\mu''\nu''}(t_1) }  \nn + \int_0^t dt_1
\braket{ Q^{(0)}_{\mu''\nu''}(t_1) Q^{(0)}_{\nu\mu'}(t) }\, \braket{
\rho_{\mu''\nu''}(t_1) \rho_{\mu\mu'}(t) } \nn + \int_0^t dt_1
\braket{Q^{(0)}_{\mu'\mu}(t) Q^{(0)}_{\mu''\nu''}(t_1) }\, \braket{
\rho_{\mu'\nu}(t) \rho_{\mu''\nu''}(t_1) }  \nn - \int_0^t dt_1
\braket{ Q^{(0)}_{\mu''\nu''}(t_1) Q^{(0)}_{\mu'\mu}(t) }\, \braket{
\rho_{\mu''\nu''}(t_1) \rho_{\mu'\nu}(t) }. \ea
An operator $Q^{(0)}_{\mu\nu}(t)$ is defined by Eq.~(\ref{Q1}) where
the stochastic phases $\xi_p$ are replaced by their unperturbed
values $\xi_p^{(0)} = \sum_k z_{kp}\, p_k^{(0)}$ that have free
Heisenberg operators $p_k^{(0)}$ of the bath. A time evolution of
the free bath operators is determined by the Hamiltonian $H_B$
(\ref{HB}). We also assume that there are sums over repeated indices
$\mu', \mu'', \nu''$ in the right-hand side of Eq.~(\ref{NME}). The
free bath variables  $Q^{(0)}_{\mu\nu}(t)$ are nonlinear functions
of the Gaussian bath operators $\xi_p^{(0)}$. Therefore,
non-Markovian equations (\ref{NME}) are valid at the small
system-bath coupling only. We consider a regime of incoherent
tunneling for the probe qubit. This regime takes place at the small
tunneling amplitude $\Delta_p$. It follows from Eq.~(\ref{Q1}) that
the small $\Delta_p$ is related to the small system-bath
interaction. Within the perturbation theory in terms of the
parameter $\Delta_p$ we assume that in Eq.~(\ref{NME}) the
correlation functions of the qubit operators, such as $\braket{
\rho_{\mu\mu'}(t) \rho_{\mu''\nu''}(t_1) }$, can be calculated using
the free evolution equations. It is convenient to reduce the
operator $\rho_{\mu''\nu''}(t_1)$ to the operator
$\rho_{\mu''\nu''}(t)$ in such a way that
  $$\rho_{\mu''\nu''}(t_1) = e^{- i\,
\omega_{\mu''\nu''} (t-t_1) }\, \rho_{\mu''\nu''}(t),$$ so that  the
correlator is given by the equation $$\braket{ \rho_{\mu\mu'}(t)
\rho_{\mu''\nu''}(t_1) } = \delta_{\mu'\mu''}\, e^{- i
\omega_{\mu''\nu''} (t-t_1) }\,\rho_{\mu\nu''}(t),$$ where
$\delta_{\mu'\mu''}$ is the Kronecker delta.

 A diagonal element
$P_{\mu} = \langle \rho_{\mu\mu}\rangle $ of the averaged matrix
$\langle \rho_{\mu\nu}\rangle$ defines the probability to find the
source-probe system in the state $\ket{\Psi_{\mu}}$. It follows from
Eqs.~(\ref{NME}) that these probabilities are governed by  the set
of  master equations,
\ba \label{P1} \dot P_{\mu} + \Gamma_{\mu} P_{\mu} = \sum_{\nu}
\Gamma_{\mu\nu} P_{\nu}. \ea
Here $\Gamma_{\mu} = \sum_{\nu} \Gamma_{\nu\mu}$ is a relaxation
rate and $\Gamma_{\mu\nu}$ is a relaxation matrix defined by the
equation
\ba \label{Gam2} \Gamma_{\mu\nu} = \int_0^t d t_1
\braket{Q^{(0)}_{\nu\mu}(t) Q^{(0)}_{\mu\nu}(t_1) }\, e^{- i
\omega_{\mu\nu} (t-t_1) } + h.c. \ea
The bath operator $Q^{(0)}_{\mu\nu}(t)$ is given by Eq.~(\ref{Q1})
where $\xi_p^{(0)}$ is a free Gaussian operator,
$\xi_p^{(0)}~=~\sum_k z_{kp}\, p_k^{(0)}$. The commutator and the
correlation function of these Heisenberg operators taken at
different moments of time are determined by the following
expressions
\ba \frac{1}{2}\, [ \xi_p^{(0)}(t), \xi_p^{(0)}(t') ]_{-} = - i
\sum_k \frac{m_k \omega_k z_{kp}^2}{2} \, \sin \omega_k (t - t') =
\nn  - i \int \frac{d \omega}{2 \pi} \, \frac{
\chi''_p(\omega)}{\omega^2} \sin \omega (t-t') , \nn \left<
\frac{1}{2}\, [ \xi_p^{(0)}(t), \xi_p^{(0)}(t') ]_{+}\right> = \nn
\sum_k \frac{m_k \omega_k z_{kp}^2}{2} \,\coth\left(
\frac{\omega_k}{2 T} \right)\, \cos\omega_k (t - t') = \nn \int
\frac{d \omega}{2 \pi} \, \frac{ S_p(\omega)}{\omega^2} \cos \omega
(t-t') , \hspace{1cm} \ea
with $T$ being the equilibrium temperature of the free bath. The
dissipative properties of the bath are defined by the imaginary part
of its susceptibility  $\chi''_p(\omega)$ and by the spectrum
$S_p(\omega)$. They are described by the following formulas
\ba \chi''_p(\omega) = \pi \sum_k \frac{m_k \omega_k^3 z_{kp}^2}{2}
\, [ \delta (\omega - \omega_k) - \delta(\omega + \omega_k) ], \nn
 S_p(\omega) = \chi''_p(\omega) \,\coth\left(
\frac{\omega}{2 T} \right) = \nn \pi \sum_k \frac{m_k \omega_k^3
z_{kp}^2}{2} \,\coth\left( \frac{\omega_k}{2 T} \right)\times \nn
\left[ \delta (\omega - \omega_k) + \delta(\omega + \omega_k)
\right].\hspace{1cm}  \ea
For the correlator of free variables $Q_{\mu\nu}^{(0)}$ of the bath
we obtain
\ba  \braket{Q^{(0)}_{\nu\mu}(t) Q^{(0)}_{\mu\nu}(t') } =
\Delta_{\mu\nu}^2 \, \Phi_p(t-t'), \ea
where the prefactor $\Delta_{\mu\nu}^2$ is defined as
\ba \label{DeL1} \Delta_{\mu\nu}^2 = \Delta_p^2\, ( \, | \langle
\Psi_{\mu}|\downarrow_p\rangle\langle \uparrow_p | \Psi_{\nu}
\rangle |^2 + \nn | \langle \Psi_{\mu}|\uparrow_p\rangle\langle
\downarrow_p | \Psi_{\nu} \rangle |^2 \, ). \ea
The characteristic functional of the bath is given by the formula:
\ba \Phi_p (t-t') = \left< e^{2 i \xi_p^{(0)}(t)}\,  e^{ - 2 i
\xi_p^{(0)}(t')}\right> = \nn \exp \left[ - \,4 \int \frac{d\omega}{
2 \pi} \,S_p(\omega) \, \frac{ 1 - \cos \omega (t - t') }{\omega^2}
- \right. \nn \left. 4 \,i \, \int \frac{d \omega}{ 2 \pi}
\,\chi''_p(\omega)\, \frac{\sin \omega (t - t')}{\omega^2} \right].
\ea
The heat bath acting on the probe qubit may have both, low-frequency
and high-frequency, components \cite{Lanting11}. In this case the
dissipative function $\chi''_p(\omega)$ is represented as a sum of
the low-frequency susceptibility, $\chi''_{LF}(\omega)$, and the
high-frequency function $\chi''_{HF}(\omega)$:
$\chi''_p(\omega)~=~\chi''_{LF}(\omega) + \chi''_{HF}(\omega).$ The
functional $\Phi_p(t,t')$ is equal to the product of the
low-frequency and high-frequency parts: $\Phi_p(\tau) =
\Phi_{LF}(\tau) \, \Phi_{HF}(\tau). $ Here, the low-frequency factor
is determined by the formula
\ba \Phi_{LF}(\tau) = e^{- i \epsilon_p\, \tau} \, \exp \left(-
\frac{W^2 \tau^2}{2}\right), \ea
with the reorganization energy $\epsilon_p$ and with the width $W$
defined by the following equations,
\ba \epsilon_p = 4 \int \frac{d \omega}{2 \pi} \,
\frac{\chi''_{LF}(\omega)}{\omega}, \nn  W^2 = 4 \int
\frac{d\omega}{2 \pi}\, S_{LF}(\omega) = 2 T \epsilon_p. \ea
The high-frequency noise acting on the probe qubit is usually
described by the Ohmic spectral density $$\chi''_{HF}(\omega) =
\eta\, \omega \,e^{-|\omega|/\omega_c},$$ where $\eta$ is a small
dimensionless coupling constant and $\omega_c$ is the cutoff
frequency \cite{Leggett87}. In this case the high-frequency factor
$\Phi_{HF}(\tau)$ of the functional $\Phi_p(\tau)$ is given by the
expression
\ba \Phi_{HF}(\tau) = \left[ \frac{1}{1 + i \omega_c \tau} \,
\frac{\pi T \tau}{\sinh(\pi T \tau)} \right]^{4 \eta/\pi}. \ea
The relaxation matrix (\ref{Gam2}) can be written as
\ba \label{Gam3} \Gamma_{\mu\nu} = \Delta_{\mu\nu}^2 \;
\int_0^\infty d \tau \;e^{- i (\omega_{\mu\nu} + \epsilon_p) \tau}
\; e^{- W^2 \tau^2/2} \times \nn \left[ \frac{1}{1 + i \omega_c
\tau} \, \frac{\pi T \tau}{\sinh(\pi T \tau)} \right]^{4 \eta/\pi} +
h.c. \ea
A more comprehensive description of the dissipative dynamics of the
open quantum system has been carried out in Ref.~\cite{Boixo16}. We
notice that in the case of the very weak coupling of the slow probe
qubit to the high-frequency bath, when $ 4 \eta/\pi \ll 1$, the
relaxation matrix is given by the Marcus formula
\cite{AminAverin08}, 
\ba \label{Gam1} \Gamma_{\mu\nu} = \Delta_{\mu\nu}^2 \,
\sqrt{\frac{2 \pi}{ W^2}}\; \exp \left[ - \frac{(\omega_{\mu\nu} +
\epsilon_p)^2 }{2\, W^2 } \right]. \ea
For indices $\mu$ and $\nu$ we have two possible cases:
$$ (a) \;\; \ket{\Psi_{\mu} }= \ket{\Psi_n^\uparrow}\otimes
\ket{\uparrow_p},\; \ket{\Psi_{\nu}} =
\ket{\Psi_m^\downarrow}\otimes \ket{\downarrow_p},$$ and
$$ (b) \;\; \ket{\Psi_{\mu}} = \ket{\Psi_m^\downarrow}\otimes
\ket{\downarrow_p},\; \ket{\Psi_{\nu}} =
\ket{\Psi_n^\uparrow}\otimes \ket{\uparrow_p}.$$ These cases
correspond to two sets of eigensenergies and frequencies:
$$ (a) \;\; E_{\mu} = E_n^\uparrow, \; E_{\nu} = E_m^\downarrow + \epsilon, \;
\omega_{\mu\nu} = E_n^\uparrow - E_m^\downarrow - \epsilon, $$ and
$$ (b) \;\; E_{\mu} = E_m^\downarrow + \epsilon, \; E_{\nu} = E_n^\uparrow, \;
\omega_{\mu\nu} = E_m^\downarrow - E_n^\uparrow + \epsilon. $$
For these two sets we obtain the following relaxation matrices:
\ba \label{Gam2} (a) \;\; \Gamma_{\mu\nu}^{(a)} = \Delta_p^2\,
|\langle \Psi_n^\uparrow | \Psi_m^\downarrow \rangle |^2 \,\times
\nn \sqrt{\frac{2 \pi}{ W^2}}\; \exp \left[ - \frac{(E_n^\uparrow -
E_m^\downarrow - \epsilon + \epsilon_p)^2 }{2\, W^2 } \right], \nn
(b) \;\; \Gamma_{\mu\nu}^{(b)} = \Delta_p^2\, |\langle
\Psi_n^\uparrow | \Psi_m^\downarrow \rangle |^2 \times \nn
\sqrt{\frac{2 \pi}{ W^2}}\; \exp \left[ - \frac{(E_n^\uparrow -
E_m^\downarrow - \epsilon - \epsilon_p)^2 }{2\, W^2 } \right]. \ea
If we start the QTS experiment with the probe qubit being in its
$\ket{\downarrow_p}$  state and allow the qubit to tunnel into the
$\ket{\uparrow_p}$ state, the situation is described by the master
equation (\ref{P1}) where $P_{\mu} \equiv P_m $ is the probability
to find the system in  state $\ket{\Psi_{\mu}} =
\ket{\Psi_m^\downarrow}\otimes \ket{\downarrow_p}$. The system
tunnels into the state $\ket{\Psi_{\nu}} =
\ket{\Psi_n^\uparrow}\otimes \ket{\uparrow_p}$, so that here we have
the case (b) described by the relaxation matrix
$\Gamma_{\mu\nu}^{(b)} \equiv \Gamma_{mn}$ defined in
Eq.~(\ref{Gam2}). 
 The probability
 to find the system in the state
$\ket{\Psi_{\nu}}$ is given by the variable $P_{\nu} \equiv P_n$. It
follows from Eq.~(\ref{P1}) that the escape rate $\Gamma_{\mu}$ is
determined by the transposed matrix $ \Gamma_{\nu\mu}$, since
$\Gamma_{\mu} = \sum_{\nu} \Gamma_{\nu\mu}$. In our case the matrix
$\Gamma_{\nu\mu}$ is described by the case $(a)$, so that the
relaxation rate $\Gamma_{\mu}$ is given by the formula
\ba \label{Gam4} \Gamma_{\mu} \equiv \Gamma_m = \Delta_p^2\, \sum_n
|\langle \Psi_n^\uparrow | \Psi_m^\downarrow \rangle |^2 \,\times
\nn \sqrt{\frac{2 \pi}{ W^2}}\; \exp \left[ - \frac{(E_n^\uparrow -
E_m^\downarrow - \epsilon + \epsilon_p)^2 }{2\, W^2 } \right]. \ea
As a result, the time evolution of the probability $P_m$ to find the
source-probe system in the state $\ket{\Psi_m^\downarrow}\otimes
\ket{\downarrow_p}$ is governed by the equation
\ba \label{P3} \dot P_m + \Gamma_m P_m = \sum_n \Gamma_{mn} P_n, \ea
where $P_n$ is the probability for the system to be in state
$\ket{\Psi_n^\uparrow}\otimes \ket{\uparrow_p}$.

\end{document}